\documentclass{jetpl}
\twocolumn
\title{Phonons in magnon superfluid and symmetry breaking field}

\rtitle{Phonons in magnon superfluid and symmetry breaking field}

\sodtitle{Phonons in magnon superfluid and symmetry breaking field}

\author{
G.\,E.\,Volovik\/\thanks{e-mail: volovik@boojum.hut.fi}}

\rauthor{G.\,E.\,Volovik}

\sodauthor{Volovik}

\address{ Low Temperature Laboratory, Helsinki University of
Technology,
P.O.Box 5100, FIN-02015, HUT, Finland\\~\\
Landau Institute for Theoretical Physics RAS, Kosygina 2,
119334 Moscow, Russia}

\dates{23 April 2008}{*}

\abstract{Recent experiments \cite{PhononMass,Skyba} which measured the spectrum of the Goldstone collective mode of coherently precessing state in $^3$He-B are discussed using the presentation of the coherent spin 
precession in terms of the Bose-Einstein condensation of magnons. The mass in the spectrum of  the Goldstone boson -- phonon in the superfluid magnon liquid -- is induced by the symmetry breaking field, which is played by the RF magnetic field.}

\begin{document}

\maketitle


The phase coherent precession of magnetization has been discovered in $^3$He-B in 1984 
\cite{HPDExp,HPDTheory}. Recently two experiments with homogeneously  precessing domain (HPD)  \cite{PhononMass,Skyba}  reported the gap in the spectrum of the collective mode of the coherent precession. The gap is proportional to $H_{\rm RF}^{1/2}$, where $H_{\rm RF}$ is applied RF field. Here we discuss this phenomenon using the description of the coherent precession  in terms of the Bose-Einstein condensation (BEC) of magnons \cite{Volovik20,BunkovVolovik2008}.

The hydrodynamic energy functional describing the magnon BEC in $^3$He-B has the same structure  as in any other superfluid system:
\begin{equation}
F= \frac{1}{2}\rho_{{\rm s}ij}(n)v_{\rm s}^i v_{\rm s}^j  +\epsilon(n) -\mu n + F_{\rm sb}(\alpha,n)~.
\label{HydrodynamicEnergy}
\end{equation}
where the magnon number density  $n$ and the phase of magnon Bose condensate  $\alpha$ are canonically conjugated variables; $\alpha$ is equal to the phase of the coherent precession of magnetization; ${\bf v}_{\rm s}$ is the superfluid velocity of BEC of magnons; $\mu$ is the chemical potential;  and $F_{\rm sb}$ is the symmetry breaking term which depends explicitly on $\alpha$.
The energy $F$ has some peculiar properties for coherent precession in general and for $^3$He-B in particular \cite{FominLT19}.

(i) In the coherent precession, the chemical potential $\mu$ (which is thermodynamically conjugated to magnon density $n$) is determined by frequencies: $\mu =\hbar(\omega - \omega_L)$, where $\omega$ is precession frequency; $\omega_L=\gamma H$ is Larmor frequency;  ${\bf H}=H\hat{\bf z}$ is an external magnetic field;   and $\gamma$ is gyromagnetic ratio of $^3$He atom.

(ii)  The magnon number density is related to the tilting angle $\beta$ of precession: $n=S(1-\cos\beta)/\hbar$, where $S$ is the equilibrium spin density $S=\chi H/\gamma$, and $\chi$ is magnetic susceptibility of $^3$He-B.

(iii) Magnons  typically have anisotropic mass.  In superfluid  $^3$He-B,   longitudinal and transverse masses depend on $\beta$ (and thus on $n$)  \cite{Volovik20}: 
\begin{equation}
 \epsilon({\bf p})=\hbar \omega_L   +\frac{1}{2}\left(m^{-1}\right)^{ij}p_i p_j= \hbar\omega_L+\frac{p_z^2}{2m_\parallel(\beta)} +
\frac{p_\perp^2}{2m_\perp(\beta)}~,
\label{MagnonSpectrum2}
\end{equation}
\begin{eqnarray}
\frac{1}{m_\parallel(\beta)}=2
 \frac{c_\parallel^2 \cos\beta+ c_\perp^2(1-\cos\beta)}{\hbar\omega_L} 
,
\label{LongMass}
\\
\frac{1}{m_\perp(\beta)}=\frac{c_\parallel^2(1+\cos\beta)
+ c_\perp^2(1-\cos\beta)}{\hbar\omega_L}   ,
\label{TransMass}
\end{eqnarray}
where the parameters $c_\parallel$ and $c_\perp$ are  spin wave velocities in $^3$He-B.

(iv) Anisotropic mass of magnon leads to anisotropic superfluid density and  superfluid velocity in the magnon BEC: 
\begin{equation}
\rho_{{\rm s}ij}(n)=nm_{ij} ~~,~~ v_{\rm s}^i =\hbar \left(m^{-1}\right)^{ij}  \nabla_j \alpha ~.
\label{SuperfluidDensity}
\end{equation}   
(Note that magnon superfluid velocity ${\bf v}_{\rm s}$ has nothing to do with the superfluid velocity of the background $^3$He-B liquid: the latter is ${\bf v}_{\rm s}=(\hbar/2m_3)\nabla\phi$, where $m_3$ is the mass of $^3$He atom, and $\phi$ is the phase of the order parameter in $^3$He-B. Magnon mass $m_{ij}$ is  typically much smaller than $m_3$.)

The mass supercurrent transferred by magnon liquid is isotropic:
\begin{equation}
j_i= \frac{dF}{dv_{\rm s}^i} =\hbar n \nabla_i\alpha ~,
\label{MassCurrent}
\end{equation}   
while the spin supercurrent is not:
\begin{equation}
j^i_{\rm spin}= \hbar  \left(m^{-1}\right)^{ij} j_i= \hbar n v_{\rm s}^i ~.
\label{MassCurrent}
\end{equation}  
This equation, which demonstrates that the nondissipative spin current ${\bf j}_{\rm spin}$ is transferred by the mass-current  velocity ${\bf v}_{\rm s}$ of magnon superfluid, is typical for the fully spin-polarized superfluids. The same occurs in superfluid  A$_1$ phase of $^3$He,  where only single spin component is superfluid \cite{VollhardtWolfle}.  Our magnon superfluid is also fully polarized, since all the magnons have the same direction of spin. 

(v) The magnon interaction energy $\epsilon(n)$  is provided by spin-orbit (dipole-dipole) interaction, which has a very peculiar form in $^3$He-B:
 \begin{equation}
\epsilon(n) =\frac{8\chi \Omega_L^2}{15\gamma^2} \left(
\frac{\hbar n}{S}-\frac{5}{4}\right)^2 \Theta \left(
\frac{\hbar n}{S}-\frac{5}{4}\right)~,
     \label{FHPD}
  \end{equation}
  where $ \Theta(x)$ is Heaviside step function; $\Omega_L$ is Leggett frequency (we assume that  $\Omega_L\ll \omega_L$).
  Stable coherent precession (HPD)  occurs only at large enough magnon density $n>5S/4\hbar$,  where $d^2\epsilon/dn^2>0$. This corresponds to $\cos\beta<-1/4$.
  
(vi)   Finally, the symmetry-breaking term $F_{\rm sb}(\alpha,n) = -\gamma {\bf H}_{\rm RF}\cdot{\bf S}$ is induced by the RF field ${\bf H}_{\rm RF}$, which is transverse to the applied constant field ${\bf H}$ and serves as a source of magnons. In continuous wave NMR experiments the RF field prescribes the frequency of precession, $\omega=\omega_{\rm RF}$, and thus fixes the chemical potential $\mu$; while in the state of free precession the chemical potential $\mu$ is determined by the number of pumped magnons. $F_{\rm sb}(\alpha,n)$  depends explicitly on the phase of precession $\alpha$ with respect to the direction of the RF-field in the precessing frame:
\begin{equation}
F_{\rm sb}= -  \gamma H_{\rm RF} S_\perp \cos\alpha \approx - \gamma H_{\rm RF} S \sin\beta \left(1-\frac{\alpha^2}{2}  \right) ~.
\label{SymmetryBreakingRF}
\end{equation}
In terms of the macroscopic wave function of  magnon Bose condensate $\psi=n^{1/2}e^{i\alpha}$ it has the form:
\begin{equation}
F_{\rm sb}(\psi)= - \frac{1}{2}\eta\left(\psi +\psi^*\right) ~,
\label{SymmetryBreakingRF2}
\end{equation}
where the symmetry-breaking field $\eta$ is
\begin{equation}
\eta=  \hbar\gamma H_{\rm RF}  \sqrt{\frac{2S}{\hbar}- n} ~.
\label{SymmetryBreakingField}
\end{equation}

The hydrodynamic energy in terms of the canonically conjugated hydrodynamic variables $n$ and   $\alpha$ allows us to find the spectrum the low frequency collective modes of the coherent precession. In the absence of the RF field there is a Goldstone mode coming from the $U(1)$ degeneracy  of the precessing states with respect to the condensate phase $\alpha$. This is  the sound wave (phonon) in magnon superfluid.  The sound wave velocity  is determined by compressibility of magnon superfluid and by magnon mass. Since the mass is anisotropic the phonon spectrum is also  anisotropic:
\begin{equation}
\left(c_s^2\right)^{ij}=\left(m^{-1}\right)^{ij}\frac{dP}{dn}=  n \frac{d^2\epsilon}{dn^2}  \left(m^{-1}\right)^{ij}~.
\label{SpeedSound}
\end{equation}
In the typical experiments with HPD, $\cos\beta$  is close to $-$1/4, i.e. $\cos\beta=-1/4-0$. For such  $\beta$ one has:  
 \begin{equation}
c_{s\parallel}^2= \frac{n}{m_{\parallel}}\frac{d^2E_{\rm so}}{dn^2}=\frac{2}{3}\frac{\Omega_L^2}{\omega_L^2}\left(5c_\perp^2-c_\parallel^2\right)~,
\label{SpeedSoundParallel}
\end{equation}
 \begin{equation}
c_{s\perp}^2= \frac{n}{m_{\perp}}\frac{d^2E_{\rm so}}{dn^2}=\frac{1}{3}\frac{\Omega_L^2}{\omega_L^2}\left(5c_\perp^2+3c_\parallel^2\right)~.
\label{SpeedSoundPerp}
\end{equation}
The sound in magnon subsystem propagating along the field ${\bf H}$ with the speed in Eq.(\ref{SpeedSoundParallel}) has been calculated in Ref.\cite{FominSound}  and observed in Ref. \cite{5}.

 The Goldstone boson (phonon) acquires mass (gap) due to the transverse RF field ${\bf H}_{\rm RF}$. The latter plays the role of the symmetry breaking field, since it violates the $U(1)$ symmetry of precession. The symmetry breaking $\alpha^2$ term  in Eq.~(\ref{SymmetryBreakingRF}) adds the isotropic mass (gap) to the phonon spectrum. For $\cos\beta=-1/4$ one obtains:
\begin{equation}
\omega^2_s({\bf k})= \left(c_s^2\right)^{ij} k_i k_j + m_s^2~~,~~ m_s^2=\frac{4}{\sqrt{15}} \gamma H_{\rm RF} \frac{\Omega_L^2}{\omega_L}~.
\label{SoundMode}
\end{equation}
This gap $m_s$ has been first calculated in Ref. \cite{PhononMass} using general Leggett equations for spin dynamics in $^3$He-B. It was measured in Refs. \cite{PhononMass,Skyba}.

I thank V.V. Dmitriev and V.B.  Eltsov for discussion. This work was supported in part by the Russian
Foundation for Basic Research (grant 06-02-16002-a).

\end{document}